\documentclass[aps,preprint,preprintnumbers,nofootinbib,showpacs,superscriptaddress]{revtex4}
\usepackage{graphicx,color,amsmath}
\begin{document}
\title{Study of $B\to p\bar{p}K^*$ and $B\to p\bar{p}\rho$}
\author{C. Q. Geng}
\author{Y. K. Hsiao}
\affiliation{Department of Physics, National Tsing Hua University, Hsinchu, Taiwan 300}
\affiliation{Theory group, TRIUMF, 4004 Wesbrook Mall, Vancouver, B.C., Canada}
\author{J. N. Ng}
\affiliation{Theory group, TRIUMF, 4004 Wesbrook Mall, Vancouver, B.C., Canada}
\date{\today}
\begin{abstract}
We study the three-body baryonic $B$ decays of $B\to
p\bar{p}(K^{*},\rho)$ in the standard model. The baryonic matrix
elements are calculated in terms of the $SU(3)$ flavor
symmetry and the QCD power counting rules within the the
perturbative QCD. We find that the decay branching ratios,
angular and direct CP asymmetries of ($B^{-}\to p\bar{p}K^{*-},
\bar{B}^{0}\to p\bar{p}K^{*0}, B^{-}\to p\bar{p}\rho^{-}$) are
around $(6,1,30)\times 10^{-6}$, $(13,-27,11)\%$ and
$(22,1,-3)\%$, which are consistent with the current BaBar and
Belle data, respectively. The large values of the branching ratio
in $B^{-}\to p\bar{p}\rho^{-}$ and the direct CP asymmetry in
$B^{\pm}\to p\bar{p}K^{*\pm}$ are useful to test the standard
model and search for new physics.

\end{abstract}


\maketitle
\newpage
\section{introduction}
The three-body baryonic $B$ decays of
$\bar{B}^{0}\to \Lambda
\bar{p}\pi^{+}$,
$B^{-}\to  \Lambda
\bar{p}\gamma$, $B\to p\bar{p} K$, $B^{-}\to p\bar{p} \pi^{-}$ and $B^{-}\to\Lambda\bar{\Lambda}K^{-}$
have been observed by BaBar and Belle Collaborations
at the levels of $10^{-6}$ on the decay
branching ratios
\cite{Lambdappi,Lambdapgamma,LambdaLambdaK,BelleAD,ExptBaBar,ExptBelle}
with near threshold enhancements in the dibaryon invariant mass spectra.
On the other hand, it is still inconclusive whether similar
threshold enhancements exist in the modes with the vector mesons
such as $B\to p\bar{p} K^{*}$ due to the large experimental
uncertainties in the present data,
 given by \cite{ExptBelle,BabarThesis}
\begin{eqnarray}\label{data}
Br(B^-\to p\bar p K^{*-})&=&(10.3^{+3.6+1.3}_{-2.8-1.7})\times 10^{-6}\;\text{(Belle)}\,,\nonumber\\
                               &=&(4.94\pm 1.66\pm 1.00)\times 10^{-6}\;\text{(Babar)}\,,\nonumber\\
Br(\bar B^0\to p\bar p \bar K^{*0})&<&7.6\times 10^{-6},\;\text{90\%\,C.L.}\;\text{(Belle)}\,,\nonumber\\
                               &=&(1.28\pm 0.56^{+0.18}_{-0.17})\times 10^{-6}\;\text{(Babar)}\,.
\end{eqnarray}

Theoretically, the dibaryon threshold enhancements in the three-body baryonic $B$ decays were first conjectured in Ref. \cite{HouSoni}.
Subsequently, various interpretations,
including models with a baryon-antibaryon bound state or baryonium
\cite{Baryonium}, exotic glueball states \cite{Glueball,Rosner},
fragmentation \cite{Rosner} and final
state interactions \cite{FSI} have been proposed.
Nevertheless,
 only the approachs of
the pole model \cite{HYpole,HYradi2,HYreview} and the QCD counting
rules within the framework of the perturbative QCD (PQCD)
\cite{Brodsky1,Brodsky2,Brodsky3}
lead to
consistent and systematic understandings on
$B\to {\bf B}\bar{\bf B}'M\ ({\bf
B,B'}=p,\Lambda$ and $M=D,K,\pi,\gamma)$
\cite{HYpole,HYradi2,HYreview,ChuaHouTsai,ChuaHouTsai2,ChuaHou,Tsai,Geng,AngdisppK,CP_ppKstar}.
However, for the vector meson mode of $B^-\to p\bar p K^{*-}$,
the pole model predicts that
a branching ratio
smaller than  $Br(B^-\to p\bar p K^{-})$, which does
not agree with the data as pointed out in Ref. \cite{HYreview},
while the corresponding PQCD study
has not been done in the literature yet.
In this paper, we give a systematic analysis on
$B\to p\bar{p}M_{V}$ with $M_{V}=K^{*}$ and $\rho$
 based
on the $B\to p\bar{p}$ transition hadronic matrix elements parametrized in Ref. \cite{AngdisppK}.

The  paper is organized as followed.
In Sec. II, we present
the formulation for the decays of
$B\to p\bar{p}M_{V}$ in the standard model.
We first derive the decay amplitudes via the effective Hamiltonian and parametrize
the  hadronic matrix elements in terms of various form factors.
We then list the formulas for the decay widths as well as the angular and direct
CP asymmetries.
 In Sec. III, we show the numerical analysis and
summarize our results in Sec. IV.

\section{formulation}
 From the effective Hamiltonian at the quark level
\cite{Hamiltonian,ali},
the decay amplitude of $B\to p\bar p M_{V}$ is separated into two
parts, given by
\begin{eqnarray}\label{amp}
{\cal A}(B\to p\bar p M_{V})&=&{\cal C}(B\to p\bar p M_{V})
+{\cal T}(B\to p\bar p M_{V})\;,\nonumber\\
\nonumber\\
{\cal C}(B\to p\bar p M_{V})&=&
\frac{G_F}{\sqrt 2}\bigg\{\bigg[
V_{ub}V_{uq}^* a_2\langle p\bar p|(\bar u u)_{V-A}|0\rangle-V_{tb}V_{tq}^*
\bigg (
a_3\langle p\bar p|(\bar u u+\bar d d+\bar s s)_{V-A}|0\rangle\nonumber\\
&&+a_4\langle p\bar p|(\bar q q)_{V-A}|0\rangle+a_5\langle p\bar p|(\bar u u+\bar d d+\bar s s)_{V+A}|0\rangle
\nonumber\\
&&+\frac{a_9}{2}\langle p\bar p|(2\bar u u-\bar dd-\bar ss)_{V-A}|0\rangle\bigg)\bigg]
\langle M_{V}|(\bar q b)_{V-A}|B\rangle
\nonumber\\
&&+V_{tb}V_{tq}^*2a_6\langle p\bar p|(\bar q q)_{S+P}|0\rangle \langle M_{V}|(\bar q b)_{S-P}|B\rangle\bigg\}\;,\nonumber\\
{\cal T}(B\to p\bar p M_{V})&=&\frac{G_F}{\sqrt 2}\alpha_{M_{V}}\langle M_{V}|(\bar q q')_{V-A}|0\rangle \langle p\bar p|(\bar q' b)_{V-A}|B\rangle\;,
\end{eqnarray}
where  $G_F$ is the Fermi constant, $V_{q_iq_j}$ are the CKM
matrix elements, $(\bar{q}_iq_j)_{V\pm
A}=\bar{q}_i\gamma_\mu(1\pm\gamma_5)q_j$, $(\bar{q}_iq_j)_{S\pm P
}=\bar{q}_i(1\pm\gamma_5)q_j$ and $a_i=c^{eff}_i+c^{eff}_{i\pm
1}/N_c$ with the color number $N_c$ for $i=$odd (even) in terms of
the effective Wilson coefficients $c^{eff}_i$, defined in Refs.
\cite{Hamiltonian,ali}.
In Eq. (\ref{amp}),
$q=s(d)$ and $q'=u(d)$ correspond to $B\to p\bar p K^{*}\,(\rho)$ and
the charged (neutral) modes,  and  $\alpha_{M_{V}}$ with $M_{V}=K^{*-}$,
$\rho^-$, $\bar{K}^{*0}$ and $\rho^0$
are given by
\begin{eqnarray}\label{amp2}
\alpha_{K^{*-}}&=&V_{ub}V_{us}^* a_1-V_{tb}V_{ts}^*a_4\,,\nonumber\\
\alpha_{\rho^-}&=&V_{ub}V_{ud}^* a_1-V_{tb}V_{td}^*a_4\,,\nonumber\\
\alpha_{\bar K^{*0}}&=&-V_{tb}V_{ts}^*a_4\,,\nonumber\\
\alpha_{\rho^0}&=&V_{ub}V_{ud}^* a_2+V_{tb}V_{td}^*\left(a_4-\frac{3}{2}a_9\right)\,,
\end{eqnarray}
respectively.
We note that we have used the generalized 
factorization method with the non-factorizable effects absorbed in
$N_{c}^{eff}$.

To evaluate the amplitude in Eq. (\ref{amp}), we need to know the
$B \to M_{V}$ transition form factors and the vector-meson decay constants
 as well as the time-like baryonic and $B\to p\bar{p}$
 transition form factors due to the vector, axial-vector, scalar and pseudoscalar quark currents, denoted as
$V_\mu$, $A_\mu$, $S$ and $P$, respectively.

The mesonic $B \to M_{V}$ transitions
have the following structures:
\begin{eqnarray}
\langle M_{V}|V_{\mu}|B\rangle&=&\epsilon_{\mu\nu\alpha\beta}
\varepsilon^{\ast\nu}p_B^{\alpha}p_{M_{V}}^{\beta}\frac{2V_1}{m_{B}+m_{M_{V}}}\;,\\
\langle M_{V}|A_{\mu}|B\rangle
&=&i\bigg[\varepsilon^\ast_\mu-\frac{\varepsilon^\ast\cdot
q}{q^2}q_\mu\bigg](m_B+m_{M_{V}})A_1
 + i\frac{\varepsilon^\ast\cdot q}{q^2}q_\mu(2m_{M_{V}})A_0\nonumber\\
&-&i\bigg[(p_B+p_{M_{V}})_\mu-\frac{m^2_B-m^2_{M_{V}}}{q^2}q_\mu \bigg](\varepsilon^\ast\cdot q)\frac{A_2}{m_B+m_{M_{V}}}\;,\nonumber\\
\langle M_{V}|V_{\mu}|0\rangle&=&f_{M_{V}}m_{M_{V}} \varepsilon_{\mu}^*\,,
\end{eqnarray}
where $q=p_B-p_{M_{V}}$,
 $V_1$ and $A_{0,1,2}$ are the form factors of the
$B \to M_{V}$ transition, and
  $\varepsilon^\ast_\mu$ and $f_{M_{V}}$ are
the polarization and the decay constant of the vector meson,
respectively.

For the time-like baryonic form factors, we have
\begin{eqnarray}\label{form3}
\langle p\bar{p}|V_{\mu}|0\rangle
&=&\bar u\bigg\{F_1\gamma_\mu+\frac{F_2}{m_p+m_{\bar p}}
i\sigma_{\mu\nu}(p_{ \bar p}+p_{p})_\mu\bigg\}v\nonumber\\
&=& \bar u\bigg\{[F_1+F_2]\gamma_\mu+\frac{F_2}{m_{p}+m_{ \bar p}}
(p_{ \bar p}-p_{p})_\mu\bigg\}v\;,\nonumber\\
\langle p\bar{p}|A_\mu|0\rangle&=&\bar u\bigg\{g_A\gamma_\mu+\frac{h_A}
{m_{p}+m_{\bar p}}
(p_{\bar p}+p_{ p})_\mu\bigg\}\gamma_5 v\,,\nonumber\\
\langle p\bar{p}|S|0\rangle &=&f_S\bar uv\;,\nonumber\\
\langle p\bar{p}|P|0\rangle &=&g_P\bar u\gamma_5 v\,,
\end{eqnarray}
where $\bar u$ and $v$ are the spinors of the baryon pair, while
$F_1$, $F_2$, $g_A$, $h_A$, $f_S$ and $g_P$ are the form factors.
Based on the QCD counting rules in the PQCD approach
\cite{ChuaHou,Brodsky1,Brodsky2,Brodsky3}, they can be
parameterized as
\begin{eqnarray}
\label{F1u}
F_1&=&\frac{5}{3}G_{||}+\frac{1}{3}G_{\overline{||}}\;,\;g_A=\frac{5}{3}G_{||}-\frac{1}{3}G_{\overline{||}}\;,\;
g_P=f_S=-\frac{4F_{||}}{3}\;
\end{eqnarray}
for $\langle p\bar p|(\bar u u)_{V,A,S,P}|0\rangle$ and
\begin{eqnarray}
\label{F1d}
F_1&=&\frac{1}{3}G_{||}+\frac{2}{3}G_{\overline{||}}\;,\;
g_A=\frac{1}{3}G_{||}-\frac{2}{3}G_{\overline{||}}\;,\;
g_P=f_S=\frac{F_{||}}{3}\;
\end{eqnarray}
 for $\langle p\bar p|(\bar dd)_{V,A,S,P}|0\rangle$,
where $G_{||}$, $G_{\overline{||}}$ and $F_{||}$ are
the functions of $t$ expanded by
\begin{eqnarray}\label{Gff4}
G_{||}=\frac{C_{||}}{t^2}\left[\text{ln}(\frac{t}{\Lambda^2_0})\right]^{-\gamma}\;,\;\;
G_{\overline{||}}=\frac{C_{\overline{||}}}{t^2}\left[\text{ln}(\frac{t}{\Lambda^2_0})\right]^{-\gamma}\;,\;\;
F_{||}=\frac{D_{||}}{t^2}\left[\text{ln}(\frac{t}{\Lambda^2_0})\right]^{-\gamma}\;,\;\;
\end{eqnarray}
with $\gamma=2.148$, $\Lambda_0=300$ MeV, where the power
expansion of $1/t^2$ is due to 2 gluon propagators attaching to
valence quarks to form a baryon pair. In Eq. (\ref{form3}), $F_2$ is suppressed by
$1/(t\text{ln}[t/\Lambda_0^2])$ in comparison with $F_1$ \cite{F2,F2b} and therefore
can be safely ignored; while $h_A$ is given by
\begin{eqnarray}\label{fsha}
h_A&=&-\frac{(m_{p}+m_{\bar p})^2}{t}g_A\;
\end{eqnarray}
by equation of motion.

For $B\to p\bar{p}$ transition form factors,
the general forms are given by \cite{AngdisppK}
\begin{eqnarray}
&&\langle p\bar{p}|V_{\mu}|B\rangle=i\bar u[g_1\gamma_{\mu}+g_2i\sigma_{\mu\nu}p^\nu+
g_3p_{\mu}+g_4(p_{\bar p}+p_{p})_\mu +g_5(p_{\bar p}-p_{ p})_\mu]\gamma_5v\;,\nonumber\\
&&\langle p\bar{p}|A_{\mu}|B\rangle=i\bar u[f_1\gamma_{\mu}+f_2i\sigma_{\mu\nu}p^\nu+f_3p_{\mu}
+f_4(p_{\bar p}+p_{ p})_\mu +f_5(p_{\bar p}-p_{p})_\mu]v\;,
\label{BtoBB}
\end{eqnarray}
where $p=p_B-p_{p}-p_{\bar p}$ is the emitted four
momentum.
By using the SU(3) flavor and SU(2) spin symmetries
\cite{ChuaHouTsai2,AngdisppK}, they are composed of another set of
form factors in the QCD counting rules with the relations of
\begin{eqnarray}\label{fm1u}
g_1&=&\frac{5}{3}g_{||}-\frac{1}{3}g_{\overline{||}}\,,\,f_1=\frac{5}{3}g_{||}+\frac{1}{3}g_{\overline{||}}\,,\,
g_j=\frac{4}{3}f_{||}^j=-f_j
\end{eqnarray}
for $\langle p\bar p|(\bar u b)_{V,A}|B^-\rangle$
and
\begin{eqnarray}\label{fm1d}
&&g_1=\frac{1}{3}g_{||}-\frac{2}{3}g_{\overline{||}}\,,\,
f_1=\frac{1}{3}g_{||}+\frac{2}{3}g_{\overline{||}}\,,\,
g_j=\frac{-1}{3}f_{||}^j=-f_j\,
\end{eqnarray}
for $\langle p\bar p|(\bar d b)_{V,A}|\bar B^0\rangle$,
  where $j=2,\cdots,5$ and
$g_{||}$, $g_{\overline{||}}$ and $f_{||}^i$ are
expressed by
\begin{eqnarray}\label{fm2}
g_{||}=\frac{N_{||}}{t^3}\;,\;g_{\overline{||}}=\frac{N_{\overline{||}}}{t^3}\;,\;f_{||}^j=\frac{M_{||}^j}{t^3}\;.
\end{eqnarray}
We note that the power expansions of $g_{i},f_{j}\propto 1/t^{3}$
are due to the need of
three hard gluons in the processes
\cite{AngdisppK}.
Two of them are used to create  valence quark pairs
just  like the time-like baryonic form
factors, and the third one is responsible for
kicking the
spectator quark in the $B$ meson \cite{Tsai}.
We emphasize that the assumption of the only $t$-dependence for
$g_{i}$ and $f_{i}$ in Eqs. (\ref{fm1u}), (\ref{fm1d}) and
(\ref{fm2})
 may not be correct.
In general, they  are
also functions of the other independent variable, such as
$(p_{p}+p_{M_{V}})^{2}$ or $(p_{\bar{p}}+p_{M_{V}})^{2}$.
To verify our assumption, it is important to
measure the baryonic leptonic B decays, such as
$B^{+}\to p\bar{p}\ell^{+}\nu_{\ell}$ \cite{Tsai}.

 From Eq. (\ref{amp}),
the squared
amplitude for $B\to p\bar{p}M_{V}$ after
summing over all spins is given by
\begin{eqnarray}\label{amp3}
&&|\bar {\cal A}|^2=\frac{G_F^2}{2}|\alpha_{M_{V}}|^2
f_{M_{V}}^2\bigg\{\big[2m_B^2(4E_{p}E_{\bar p}-t-m_{M_{V}}^2)+2m_{M_{V}}^2(2t-8m_p^2+m_{M_{V}}^2)\big]g_1^2\nonumber\\
&&
+24m_{M_{V}}^2m_pm_B(E_{\bar p}-E_{p})g_1 g_2
+2m_p\big[(m_B^2-t)^2-2m_{M_V}^2(m_B^2+t)\big]g_1 g_4
\nonumber\\
&&+4m_Bm_p(m_B^2-t-2m_{M_{V}}^2)(E_{\bar p}-E_{p})g_1g_5\nonumber\\
&&+2m_{M_{V}}^2\big[2m_B^2(4E_{p}E_{\bar
p}-t-m_{M_V}^2)-m_{M_{V}}^2(4m_p^2-t-2m_{M_{V}}^2)\big]g_2^2
+4m_{M_{V}}^2m_Bt(E_{\bar p}-E_{p})g_2 g_4\nonumber\\
&&+2m_{M_{V}}^2(t-4m_p^2)(m_B^2-t-2m_{M_{V}}^2)g_2 g_5
+\frac{1}{2}t\big[(m_B^2-t)^2-4m_{M_V}^2 t\big]g_4^2\nonumber\\
&&+2m_B(m_B^2-t)t(E_{\bar p}-E_{p})g_4g_5+2t[m_B^2(E_{\bar p}-E_{p})^2+m_{M_{V}}^2(t-4m_p^2)]g_5^2\nonumber\\
&&+2[m_{M_V}^4+m_{M_V}^2(m_B^2+4m_p^2+2t)+m_B^2(4E_p E_{\bar p}-t)]f_1^2\nonumber\\
&&+12m_{M_{V}}^2m_p(m_B^2-t-2m_{M_{V}}^2)f_1 f_2
+4m_B m_p(m_B^2-t)(E_{\bar p}-E_{p})f_1f_4\nonumber\\
&&+8m_p\big[m_B^2(E_{\bar p}-E_{p})^2
+m^{2}_{M_{V}}(t-4m_p^2)\big]f_1f_5\nonumber\\
&&+2m_{M_{V}}^2\big[2m_B^2(4E_{p}E_{\bar p}
-t-m_{M_{V}}^2)+m_{M_{V}}^2(8m_p^2+t+2m_{M_{V}}^2)
\big]f_2^2\nonumber\\
&&+4m_{M_{V}}^2m_Bt(E_{\bar p}-E_{p})f_2 f_4
+2m_{M_{V}}^2(t-4m_p^2)(m_B^2-t-2m_{M_{V}}^2)f_2 f_5\nonumber\\
&&+\frac{1}{2}(t-4m_p^2)\big[
(m_B^2-t)^2-4m_{M_V}^2 t\big]f_4^2\nonumber\\
&&+2m_B(t-4m_p^2)(m_B^2-t)(E_{\bar p}-E_{p})f_4f_5
+2m_B^2(t-4m_p^2)(E_{\bar p}-E_{p})^2f_5^2\bigg\}\,,
\end{eqnarray}
where $t\equiv (p_{p}+p_{\bar p})^2$,
$E_{p(\bar p)}$ is the energy of the proton (antiproton).
We note that in Eq. (\ref{amp3}),
 the
terms related to ${\cal C}(B\to p\bar p M_{V})$ have been ignored
due to their small contributions to the branching ratios $(<1\%)$.
Replacing  the proton and anti-proton energies $E_{p(\bar
p)}$ by
  the angle of $\theta$
  between the three-momenta of the vector
meson and  the proton in the dibaryon rest
frame, $i.e.$,
\begin{eqnarray}\label{Ep}
E_{p(\bar p)}&=&\frac{m_B^2+t-m_p^2\pm
\beta_p\lambda^{1/2}_t\cos\theta}{4m_B}\;
\end{eqnarray}
where $\beta_p=(1-4m_p^2/t)^{1/2}$ and
$\lambda_t=m_B^4+m_{M_{V}}^4+t^2-2m_{M_{V}}^2 t-2m_B^2 t-2m_{M_{V}}^2 m_B^2$,
we can rewrite Eq. (\ref{amp3}) as
\begin{eqnarray}\label{amptheta1}
{\it |\bar {\cal A}|^2}=\frac{G_F^2}{2}|\alpha_{M_{V}}|^2
f_{M_{V}}^2(\rho_0 +\rho_\theta\cos\theta
+\rho_{\theta^2}\cos^2\theta )\,,
\end{eqnarray}
where
\begin{eqnarray}\label{amptheta2}
\rho_0&=&\frac{1}{2}\bigg[m_{M_V}^4+16m_p^2m_{M_V}^2-2(m_B^2-3t)m_{M_V}^2+(m_B^2-t)^2\bigg](g_1^2+f_1^2)-24m_p^2m_{M_V}^2 g_1^2\nonumber\\
&&-12m_p m_{M_V}^2(m_{M_V}^2-m_B^2+t)f_1 f_2+2[m_{M_V}^4-2(m_B^2+t)m_{M_V}^2+(m_B^2-t)^2]g_1 g_4\nonumber\\
&&+8m_p m_{M_V}^2(t-4m_p^2)f_1 f_5\nonumber\\
&&+m_{M_V}^2[16m_p^2m_{M_V}^2+(m_{M_V}^2-m_B^2)^2+t^2-2m_B^2 t](g_2^2+f_2^2)-24m_p^2m_{M_V}^4 g_2^2\nonumber\\
&&+2m_{M_V}^2(t-4m_p^2)(m_B^2-m_{M_V}^2-t)(g_2 g_5+f_2 f_5)
\nonumber\\
&&+\frac{1}{2}[m_{M_V}^4-2m_{M_V}^2(m_B^2+t)+(m_B^2-t)^2][g_4^2 t+(t-4m_p^2)f_4^2]\nonumber\\
&&+2m_{M_V}^2(t-4m_p^2)[t g_5^2 +(t-4m_p^2)f_5^2]\,,
\nonumber\\
\nonumber\\
\rho_\theta&=&\beta_p \lambda_t^{1/2}\{2m_{M_V}^2[6m_p g_1 g_2+t(g_2 g_4+f_2 f_4)]\nonumber\\
&&-(m_{M_V}^2-m_B^2+t)[2m_p(g_1 g_5+f_1 f_4-2m_pf_4 f_5)+t(g_4 g_5+f_4 f_5)]\}\,,
\nonumber\\
\nonumber\\
\rho_{\theta^2}&=&-\frac{1}{2}\beta_p^2 \lambda_t[g_1^2+(f_1-2m_p f_5)^2+2m_{M_V}^2(g_2^2+f_2^2)-t(g_5^2+f_5^2)]\,.
\end{eqnarray}
\\
The decay width $\Gamma$ of $B\to
p\bar{p}M_{V}$ is given by  \cite{eqofag}
\begin{eqnarray}\label{Gamma}
\Gamma&=&\int^{+1}_{-1}{d\,\Gamma\over d\cos\theta}d\cos\theta
\;=\;
\int^{+1}_{-1}\int^{(m_B-m_{M_{V}})^2}_{4m_p^2}\frac{\beta_p\lambda^{1/2}_t}{(8\pi
m_B)^3}|\bar {\cal A}|^2\;dt\;d\cos\theta\;.
\end{eqnarray}
 From Eq. (\ref{Gamma}), we can study the partial decay width
$d\Gamma/d\cos\theta$ as a function of $\cos\theta$, $i.e.$, the
angular distribution and define the angular asymmetry by
\cite{AngdisppK,eqofag}
\begin{eqnarray}\label{AFB}
A_{\theta}(M_{V})&\equiv&\frac{\int^{+1}_0\frac{d\Gamma}{d\cos\theta}d\cos\theta
-\int^0_{-1}\frac{d\Gamma}{d\cos\theta}
d\cos\theta}{\int^{+1}_0\frac{d\Gamma}{d\cos\theta}
d\cos\theta+\int^0_{-1}\frac{d\Gamma}{d\cos\theta} d\cos\theta}\;,
\end{eqnarray}
which is equal to $(N_+-N_-)/(N_++N_-)$, where $N_\pm$ are the
events with $\cos\theta>0$ and $\cos\theta<0$, respectively.
 From Eqs. (\ref{amptheta1}), (\ref{Gamma}) and (\ref{AFB}), we
 obtain
 \begin{eqnarray}\label{AFB1}
A_{\theta}(M_{V})&=&
{1\over \Gamma}\int^{(m_B-m_{M_{V}})^2}_{4m_p^2}\rho_{\theta}\,dt\,.
\end{eqnarray}
 We can also define the direct CP asymmetry in $B\to p\bar{p}M_{V}$
by \cite{CP_ppKstar}
\begin{eqnarray}\label{Acp}
A_{CP}(M_{V})&=&\frac{\Gamma(B^-\to p\bar p
M_{V}^-)-\Gamma(B^+\to p\bar p M_{V}^+)}{\Gamma(B^-\to p\bar p M_{V}^-)
+\Gamma(B^+\to
p\bar p M_{V}^+)}\,.
\end{eqnarray}
 From Eqs. (\ref{amp3}) and (\ref{Gamma}),
 we get
\begin{eqnarray}\label{Acp1}
A_{CP}(M_{V})&=&
\frac{|\alpha_{M_{V}}|^2-|\bar
\alpha_{M_{V}}|^2}{|\alpha_{M_{V}}|^2+|\bar \alpha_{M_{V}}|^2}\,,
\end{eqnarray}
where
$\bar\alpha_{M_{V}}$ denotes the
value of the corresponding antiparticle.

\section{Numerical analysis}
In our numerical analysis, the vector-meson decay constants of
 $f_\rho$ and $f_{K^*}$ are taken to be
 \cite{decayconst}
\begin{eqnarray}
(f_\rho,f_{K^*})=(0.205,0.217)\,\text{GeV}\,,
\end{eqnarray}
and the form factors in the $B \to M_{V}$ transition as functions of $t$ are given by \cite{MFD}
\begin{eqnarray}
V_{1}[A_{0}](t)&=&\frac{V_{1}[A_{0}](0)}{(1-t/M_{1[2]}^2)
(1-\sigma_1 t/M_{1[2]}^2+\sigma_2 t^2/M_{1[2]}^4)}\,,
\nonumber\\
A_{1,2}(t)&=&\frac{A_{1,2}(0)}{1-\sigma_1 t/M_2^2+\sigma_2 t^2/M_2^4}\,,
\end{eqnarray}
where the values of $V(0)$, $A_{0,1,2}(0)$, $\sigma_{1,2}$ and $M_{1,2}$ are shown in
Table \ref{MF} \cite{MFD}.
\begin{table}[h!]
\caption{ \sl Form factors of $B\to K^{*}\,(\rho)$ at $t=0$
in Ref. \cite{MFD} with
 $(M_1,M_2)=(5.37,5.42)$ and $(5.27,5.32)$ GeV for  $K^*$ and
 $\rho$, respectively.
}\label{MF}
\begin{tabular}{|c|c|c|c|c|}
\hline
$B\to K^{*}\,(\rho)$&$V_1$&$A_0$&$A_1$&$A_2$\\\hline
f(0)       &0.44 (0.31) &0.45 (0.30) &0.36 (0.26) &0.32 (0.24)\\
$\sigma_1$ &0.45 (0.59) &0.46 (0.54) &0.64 (0.73) &1.23 (1.40)\\
$\sigma_2$ &-----&-----&0.36 (0.10) &0.38 (0.50)\\\hline
\end{tabular}
\end{table}
\normalsize
 From the experimental data in Refs. \cite{Dstarpn,Lambdappi},
we obtain the time-like form factor related coefficients of $C_{||}$ ,
$C_{\overline{||}}$ and $D_{||}$ in Eq. (\ref{Gff4})
to be \cite{NF}
\begin{eqnarray}\label{C1}
C_{||}=93.1^{+6.4}_{-6.9}\,GeV^{4}\;,\;\;
C_{\overline{||}}=-203.8^{+103.6}_{-\;\;84.0}\,GeV^{4}\;,\;\;
D_{||}=189.9^{+13.1}_{-14.1}\,GeV^{4}\;.
\end{eqnarray}
Here, we have assumed that $C_{||,\overline{||}}$ and $D_{||}$
are real since
their imaginary parts
are expected to be
 small
\cite{dispersion}.
By fitting
the measured branching ratios of $B^- \to p\bar p
K^{(*)-}$ and $\bar B^0 \to p\bar p \bar K^{*0}$ and angular
distribution of $B^- \to p\bar p K^-$
\cite{ExptBelle,BabarThesis,BelleAD,ExptBaBar},
the values of $N_{||,\overline{||}}$ and
$M_{||}^i$ in Eq. (\ref{fm2}) are
given by \cite{AngdisppK}
\begin{eqnarray}
&&N_{||}=124.0\pm 26.5\, GeV^5,\,N_{\overline{||}}=-200.2\pm 51.1\,GeV^5,\,M_{||}^2=-247.0\pm 48.7\,GeV^4,\, \nonumber\\
&&M_{||}^4=-22.1\pm 15.7\,GeV^4,\,M_{||}^5=229.4\pm 22.7\,GeV^4\,.
\end{eqnarray}
 Note that there are no
terms related to $M_{||}^3$ ( or $g_3$ and $f_3$) due to the
relation of $\varepsilon \cdot p_{M_{V}}=0$. To compare
 the form factors  $f_{i}$ and $g_{i}$ in Eqs. (\ref{fm1u})
and  (\ref{fm1d}), we show their central values
in Table \ref{fgi}, respectively.
It is interesting to note that
\begin{eqnarray}
\label{Relation1}
f_{1}(B^-\to p\bar p)&\sim& -1.5f_{1}(\bar{B}^0\to p\bar p)\,,
\nonumber\\
g_{1}(B^-\to p\bar p)&\sim& 1.5g_{1}(\bar{B}^0\to p\bar p)\,,
\nonumber\\
f_{i}(B^-\to p\bar p)&\sim& 4f_{i}(\bar{B}^0\to p\bar p)\,,\ (i=2,4,5)\,,
\nonumber\\
g_{i}(B^-\to p\bar p)&\sim& -4g_{i}(\bar{B}^0\to p\bar p)\,,\ (i=2,4,5)\,.
\end{eqnarray}
In addition, we have
\begin{eqnarray}
\label{Relation2}
f_{2}\sim -f_{5}\gg (\ll) f_{4}\,,
\nonumber\\
g_{2}\sim -g_{5}\ll (\gg) g_{4}\,,
\end{eqnarray}
for the charged (neutral) $B$ transition.

\begin{table}[h!]
\caption{{\sl Central values of $\hat{f}_{i}(\hat{g}_{i})\equiv
t^{3} f_{i}(g_{i}) $ (in units of $GeV^{4}$).} }\label{fgi}
\begin{tabular}{|c|c|c|c|c|}
\hline
                      &$\hat{f}_1(\hat{g}_1)/m_{p}$&$\hat{f}_2(\hat{g}_2)$&$\hat{f}_4(\hat{g}_4)$&$\hat{f}_5(\hat{g}_5)$\\\hline
     $B^-\to p\bar p$&149.3(291.4)&329.3(-329.3)&29.5(-29.5)&-305.9(305.9)\\
$\bar B^0\to p\bar p$&-98.2(186.4)&-82.3,(82.3)&-7.3(7.3)&76.5(-76.5)\\\hline
\end{tabular}
\end{table}

\begin{figure}[t!]
\centering
\includegraphics[width=2.0in]{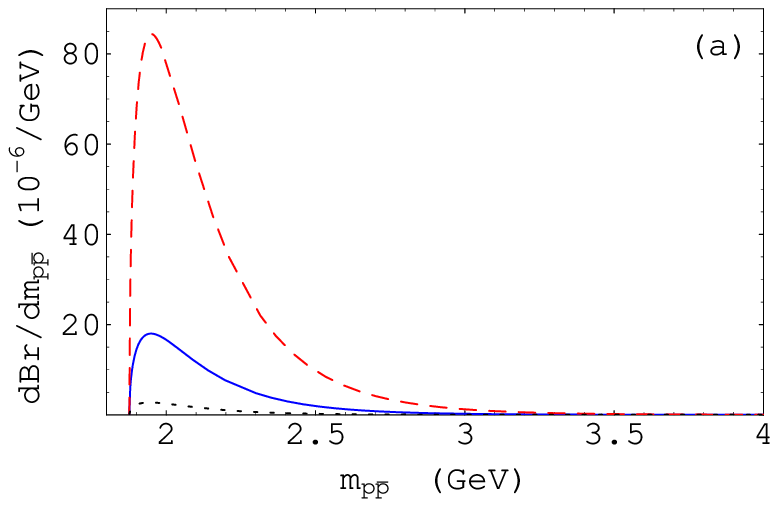}
\includegraphics[width=2.1in]{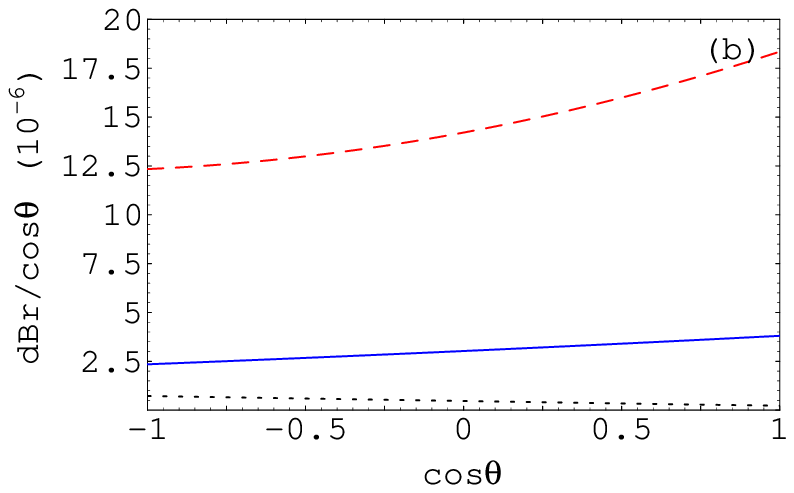}
\caption{\label{fig} Differential branching fractions of (a) 
$dBr/dm_{p\bar p}$  and (b)  
$dBr/dcos\theta$ as functions of
$m_{p\bar p}$ and $cos\theta$, where the solid, dot and dash lines
represent the decays of $B^-\to p\bar p K^{*-}$, $\bar B^0\to p\bar p K^{*0}$ and
$B^-\to p\bar p \rho^-$, respectively.}
\end{figure}

 In Figs. \ref{fig}a and \ref{fig}b,
 we show the differential branching ratios 
 of $dBr/dm_{p\bar p}$ and  
$dBr/dcos\theta$ as functions of
 $m_{p\bar p}$ and  $cos\theta$
 which demonstrate the
apparent threshold enhancement and angular asymmetry,
respectively.
 The
inetrgrated results for the decay branching ratios and the angular
and direct CP asymmetries are presented in Table \ref{table1}

\begin{table}[h!]
\caption{ \sl Decay branching ratios (in units of
$10^{-6}$) and angular  and direct CP asymmetries
in $B\to p\bar{p}M_{V}\ (M_{V}=K^{*},\rho)$, where the errors
are from the experimental data of the three-body baryonic $B$ decays.
}\label{table1}
\begin{tabular}{|c|c|c|c|}
\hline
$M_{V}$ &Br&$A_\theta(M_{V})$&$A_{CP}(M_{V})$\\
\hline
$K^{*\pm}$  &$6.0\pm1.3$ &$0.13\pm 0.05$& $0.22\pm 0.01$\\
$ K^{*0}$   &$0.9\pm 0.3$  &$-0.27\pm 0.06$& $0.013\pm0.001$ \\
$\rho^{\pm}$&$28.8\pm 2.1$ &$0.11\pm 0.06$& $-0.029\pm 0.009$\\
\hline
\end{tabular}
\end{table}
Here we have averaged the particle and antiparticle contributions
for the branching ratios and angular asymmetries.
We note that in the table we have only included the errors
 from
 the experimental data of the three-body baryonic $B$ decays,
  while the others such as
 those from nonfactorizable effects and CKM matrix elements can be referred to Ref. \cite{CP_ppKstar}. In addition, we have ignored 
 the imaginary parts of the form factors in our determinations of
 the direct CP asymmetries.

From Table \ref{table1}, we see that $Br(\bar B^0\to p\bar p \bar
K^{*0})=0.9\pm 0.3$ is consistent with the experimental value in
Eq. (\ref{data}). In addition, we can explain the inequality of
$Br(B^{-}\to p\bar{p}K^{*-})>Br(\bar{B}^{0}\to p\bar{p}K^{*0})$ in
terms of the QCD counting rules illustrated in Eq.
(\ref{Relation1}), which is also demonstrated in Fig. \ref{fig}a.
The large branching ratio of $Br(B^-\to p\bar p \rho^-)\simeq
30\times 10^{-6}$ can be understood from the ratio of
$|V_{ub}V_{ud}^*a_1 f_{\rho}/V_{tb}V_{ts}^*a_4 f_{K*}|^2\simeq 4$
as the baryonic $B\to p\bar p$ transition form factors are the
same between $B^-\to p\bar p \rho^-$ and $B^-\to p\bar p K^{*-}$.
It is interesting to point out that if $Br(B^-\to p\bar p
K^{*-})\sim10\times 10^{-6}$ as indicated by the central value of
the Belle result \cite{ExptBelle} in Eq. (\ref{data}), we find
that $Br(B^-\to p\bar p \rho^-)\sim 50\times 10^{-6}$ while
$Br(\bar B^0\to p\bar p\bar K^{*0})\sim 1.2\times 10^{-6}$ which
is close to the central value of the BaBar data in Eq.
(\ref{data}).
Since the main contributions to
$\bar B^0\to p\bar p \rho^0$ are color suppressed $a_2\simeq a_1/100$ with $N_c=3$ as seen from Eqs. (\ref{amp}) and
(\ref{amp2}),
the prediction of $Br(\bar B^0\to p\bar p \rho^0)$
is in the order of $10^{-9}$.
Even if $a_2$ ends up around
the order of $10^{-1}$ with $N_c=(2,\infty)$ for
possible nonfactorizable effects, it merely
shifts to be $O(10^{-7})$.
Although currently there are no data
about the branching ratios involving $\rho$ as well as the angular
distributions,
our predictions, in particular the large decay branching ratio for
the charged $\rho$ modes,  provide good tests for the PQCD approach
 in the ongoing and future $B$ factories.

We remark that our results of
$A_\theta(
K^{*-},\rho)\sim
10\%$
and $|A_\theta( K^{*0})|\sim 30\%$
are  smaller than
$A_{\theta}( K^{-})=0.59^{+0.08}_{-0.07}$ measured by Belle \cite{BelleAD}.
It is interesting to note that
the large value of $A_{CP}(
K^{*\pm})\sim 22\%$ is in agreement with  the BABAR data of $(26\pm 19)\%$ as given in Ref.
\cite{BabarThesis}. On the other hand,
the direct CP asymmetries in $\bar B^0\to p\bar p \bar K^{*0}$ and
$B\to p\bar p\rho$ are too small to be measured.

\section{Conclusions}
We have studied the three-body baryonic $B$ decays of
$B\to p\bar p M_{V}$ in the standard model. The baryonic matrix elements
have been calculated in terms of the $SU(3)$ flavor symmetry and the QCD power counting rules within the PQCD.
Explicitly, we have found that
the decay Brs, $A_{\theta}$ and $A_{CP}$ of
($B^{-}\to p\bar{p}K^{*-}, \bar{B}^{0}\to p\bar{p}K^{*0},
 B^{-}\to p\bar{p}\rho^{-}$) are
around $(6,1,30)\times 10^{-6}$, $(13,-27,11)\%$ and
$(22,1,-3)\%$, respectively. Our result of $Br(\bar B^0\to p\bar p
K^{*0})$ is consistent with the BaBar and Belle data, while that
of $Br(B^-\to p\bar p\rho^-)$ should be observed by
 the ongoing $B$ experiments at the current B factories soon.
Finally, we remark that our large prediction of
$A_{CP}(K^{*\pm})$ can be used to test the standard model and search for new physics. More precise measurements are clearly needed at the current and future B factories.
\\

\noindent
{\bf Acknowledgments}

This work is supported in part by
the National Science Council of
R.O.C. under Contract \#s: NSC-95-2112-M-007-059-MY3
and NSC-96-2918-I-007-010 and
the Natural Science and Engineering Council of Canada.


\end{document}